\documentclass[reprint, aps, superscriptaddress, pra]{revtex4-2}
\usepackage{graphicx}
\usepackage{float}
\usepackage{amsmath}
\usepackage{amsfonts}
\usepackage{svg}
\usepackage{amssymb}
\usepackage{commath}
\usepackage{braket}
\usepackage{multirow}
\usepackage[utf8]{inputenc}
\usepackage[margin=1in]{geometry}
\usepackage{comment}
\usepackage{dcolumn}
\usepackage{bm}
\usepackage{textgreek}
\usepackage{xr-hyper}
\usepackage{hyperref}
\usepackage[capitalize]{cleveref}
\usepackage{xcolor}
\makeatletter
\newcommand*{\addFileDependency}[1]{
  \typeout{(#1)}
  \@addtofilelist{#1}
  \IfFileExists{#1}{}{\typeout{No file #1.}}
}
\makeatother

\usepackage{tikz,xcolor,hyperref}
\definecolor{lime}{HTML}{A6CE39}
\DeclareRobustCommand{\orcidicon}{
	\begin{tikzpicture}
	\draw[lime, fill=lime] (0,0) 
	circle [radius=0.16] 
	node[white] {{\fontfamily{qag}\selectfont \tiny ID}};
	\draw[white, fill=white] (-0.0625,0.095) 
	circle [radius=0.007];
	\end{tikzpicture}
	\hspace{-2mm}
}
\foreach \x in {A, ..., Z}{\expandafter\xdef\csname orcid\x\endcsname{\noexpand\href{https://orcid.org/\csname orcidauthor\x\endcsname}
			{\noexpand\orcidicon}}
}

\begin{document}

\preprint{APS/123-QED}
\title{Electrically-triggered spin-photon devices in silicon}

\author{Michael Dobinson\orcidA{}}
\affiliation{Department of Physics, Simon Fraser University, Burnaby, British Columbia, Canada}
\affiliation{Photonic Inc., Coquitlam, British Columbia, Canada}
\author{Camille Bowness}
\affiliation{Department of Physics, Simon Fraser University, Burnaby, British Columbia, Canada}
\affiliation{Photonic Inc., Coquitlam, British Columbia, Canada}
\author{Simon A. Meynell}
\affiliation{Department of Physics, Simon Fraser University, Burnaby, British Columbia, Canada}
\affiliation{Photonic Inc., Coquitlam, British Columbia, Canada}
\author{Camille Chartrand}
\affiliation{Department of Physics, Simon Fraser University, Burnaby, British Columbia, Canada}
\affiliation{Photonic Inc., Coquitlam, British Columbia, Canada}
\author{Elianor Hoffmann}
\affiliation{Département de physique, Université Paris-Saclay, 91190 Gif-sur-Yvette, France}
\author{Melanie Gascoine}
\affiliation{Department of Physics, Simon Fraser University, Burnaby, British Columbia, Canada}
\affiliation{Photonic Inc., Coquitlam, British Columbia, Canada}
\author{Iain MacGilp}
\affiliation{Photonic Inc., Coquitlam, British Columbia, Canada}
\author{Francis Afzal}
\affiliation{Photonic Inc., Coquitlam, British Columbia, Canada}
\author{Christian Dangel}
\affiliation{Photonic Inc., Coquitlam, British Columbia, Canada}
\author{Navid Jahed}
\affiliation{Photonic Inc., Coquitlam, British Columbia, Canada}
\author{Michael L. W. Thewalt\orcidC{}}
\affiliation{Department of Physics, Simon Fraser University, Burnaby, British Columbia, Canada}
\affiliation{Photonic Inc., Coquitlam, British Columbia, Canada}
\author{Stephanie Simmons}
\affiliation{Department of Physics, Simon Fraser University, Burnaby, British Columbia, Canada}
\affiliation{Photonic Inc., Coquitlam, British Columbia, Canada}
\author{Daniel B. Higginbottom\orcidB{}}
\affiliation{Department of Physics, Simon Fraser University, Burnaby, British Columbia, Canada}
\affiliation{Photonic Inc., Coquitlam, British Columbia, Canada}

\date{\today}
\begin{abstract} 
Quantum networking and computing technologies demand scalable hardware with high-speed control for large systems of quantum devices. Solid-state platforms have emerged as promising candidates, offering scalable fabrication for a wide range of qubits. Architectures based on spin-photon interfaces allow for highly-connected quantum networks over photonic links, enabling entanglement distribution for quantum networking and distributed quantum computing protocols. With the potential to address these demands, optically-active spin defects in silicon are one proposed platform for building quantum technologies. Here, we electrically excite the silicon T centre in integrated optoelectronic devices that combine nanophotonic waveguides and cavities with p-i-n diodes. We observe single-photon electroluminescence from a cavity-coupled T centre with $g^{(2)}(0)=0.05(2)$. Further, we use the electrically-triggered emission to herald the electron spin state, initializing it with 92(8)\% fidelity. This shows, for the first time, electrically-injected single-photon emission from a silicon colour centre and a new method of electrically-triggered spin initialization. These findings present a new telecommunications band light source for silicon and a highly parallel control method for T centre quantum processors, advancing the T centre as a versatile defect for scalable quantum technologies.
\end{abstract}

\maketitle

\section{Introduction}
\label{sec:introduction}
Utility-scale quantum information technologies will require developing high-speed control methods for dense networks of quantum devices~\cite{simmons2024Scalable}. The miniaturization of classical computing technologies has shown the advantages of semiconductor platforms in this regard, and a number of promising platforms for quantum technologies have emerged that combine spins in a solid-state host with an optical interface~\cite{awschalom2018Quantum}. These spin-photon interfaces (SPIs) include semiconductor quantum dots~\cite{yu2023Telecomband}, atomic defects~\cite{yin2013Opticalb}, and more broadly, colour centres in diamond~\cite{pompili2021Realization}, silicon carbide~\cite{castelletto2020Silicona}, and recently silicon~\cite{bergeron2020SiliconIntegrated, higginbottom2022Optical, johnston2024Cavitycoupled, xiong2024Computationally}. Silicon colour centres are of particular interest for quantum technologies due to the scalability of the silicon platform, their competitive optical properties, and integration with conventional photonic and electronic devices~\cite{simmons2024Scalable, higginbottom2023Memory, afzal2024Distributed}. Native silicon emitters also have applications in classical technologies where integration with silicon photonics is a key criterion, such as on-chip electrically injected light sources, single-photon sources, and lasers~\cite{yuan2002Electrically,zhou2015Onchip,zhou2023Prospects,shekhar2024Roadmapping}.

\begin{figure*}
  \makebox[\textwidth][c]{\includegraphics[width=180mm]{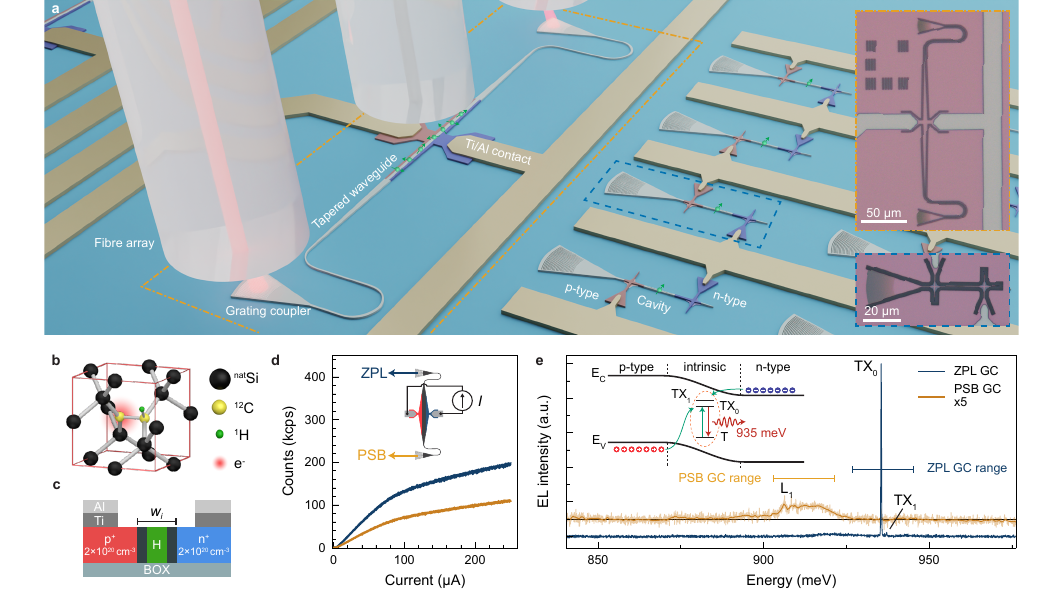}}%
  \caption{\textbf{Integrated optoelectronic devices for control of silicon T centres. a}, Illustration of the optoelectronic devices. T centres (green circles with arrows) are generated in regions within each device, defined by the hydrogen implantation mask. The insets show optical microscope images of the tapered waveguide (orange) and cavity (blue) devices. \textbf{b}, Chemical structure of the T centre. \textbf{c}, Device layer cross-section of a p-i-n diode. \textbf{d},  Electroluminescence (EL) saturation curve for a tapered waveguide device. The inset shows a schematic of electrical injection of a tapered waveguide device with a constant current. The top grating coupler (GC) is tuned for the zero-phonon line (ZPL) of the T centre and the bottom GC is tuned for the phonon sideband (PSB). \textbf{e}, EL spectrum of a T centre ensemble in a tapered waveguide device. Horizontal lines show the passband FWHM for the ZPL and PSB GCs of 18(4)~meV and 18(2)~meV, respectively. The inset shows the mechanism for EL of the T centre in a diode under forward bias.}
  \label{fig:introduction}
\end{figure*}

The silicon T centre is a candidate for quantum technologies,  possessing a valuable combination of long-lived spins and a native SPI operating in the low-loss telecommunications O-band~\cite{bergeron2020SiliconIntegrated}. Quantum operations can be performed between T centres with a broker-client scheme, using the electron spin for communication and the nuclear spin for memory~\cite{brunelle2024Silicon}. This makes the T centre an enticing solution for scalable quantum memories~\cite{higginbottom2023Memory} and distributed quantum computing~\cite{simmons2024Scalable}. A recent commercial demonstration has also shown remote entanglement between T centres~\cite{afzal2024Distributed}, placing it on the short list of remotely entangled SPIs, which also includes quantum dots~\cite{stockill2017PhaseTuned}, trapped neutral atoms~\cite{hofmann2012Heralded}, trapped ions~\cite{drmota2024Verifiable}, and diamond colour centres~\cite{pompili2021Realization, knaut2024Entanglement}. 

The optical properties of the T centre have been studied in bulk silicon, identifying it as an SPI with a bound exciton optical transition~\cite{bergeron2020SiliconIntegrated}.  The T centre has also been generated in silicon-on-insulator (SOI)~\cite{macquarrie2021Generating} and recently has been integrated into photonic waveguides~\cite{deabreu2023Waveguideintegrated, lee2023HighEfficiency} and cavities~\cite{islam2023CavityEnhanced,johnston2024Cavitycoupled,afzal2024Distributed}. First-principles calculations have predicted the charge states and formation conditions of the T centre~\cite{dhaliah2022Firstprinciples}. The T centre interacts with electric fields~\cite{clear2024Optical}, offering further opportunities for control. Other platforms, such as quantum dots and colour centres in silicon carbide and diamond have already shown capabilities using electronic devices, including Stark tuning~\cite{anderson2019Electrical, candido2021Suppression,delascasas2017Stark,ruhl2020Stark,lukin2020Spectrally}, spin initialization using spin-polarized carrier injection~\cite{cadiz2018Electrical}, electrically-injected single-photon emission~\cite{lohrmann2015Singlephoton,sato2018Room}, charge depletion for mitigating spectral diffusion~\cite{anderson2019Electrical, candido2021Suppression}, charge state tuning~\cite{widmann2019Electrical,luhmann2020ChargeState,bathen2019Electrical,kato2013Tunable,murai2018Engineering,rieger2024Fast}, and spin-charge readout~\cite{niethammer2019Coherent,siyushev2019Photoelectrical}. The G and W centres---silicon colour centres similar to the T centre---have also been integrated with electronic devices in ensembles, demonstrating electrical injection~\cite{bao2007Point,buckley2017Allsilicon,buckley2020Optimization, ebadollahi2024Fabrication} and recently, electrical manipulation of the G centre~\cite{day2024Electrical}. However, unlike the T centre, the G and W centres have no ground state electron spin~\cite{chartrand2018Highly} to operate as an SPI. 

In this work, we present two types of integrated optoelectronic T centre devices. We observe electroluminescence (EL) from ensembles of waveguide-coupled T centres. We also report single-photon EL from a single cavity-coupled T centre, with $g^{(2)}(0)=0.05(2)$. To our knowledge, this is the first demonstration of single-photon EL from a silicon colour centre. Simultaneous electrical and optical access enables a new control scheme in which we initialize the T centre's electron spin by spectrally-resolved EL. We report a state preparation and measurement (SPAM) fidelity of 92(8)\%, limited by the background luminescence from other defects. This first demonstration of spin initialization by heralded EL is a foundational component for new spin-spin entanglement techniques. These findings enable new, scalable quantum technologies based on T centre quantum processors.

\begin{figure*}
  \makebox[\textwidth][c]{\includegraphics[width=180mm]{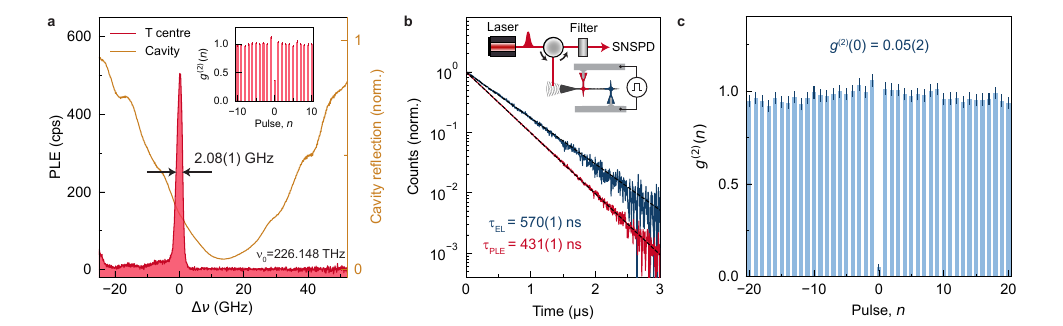}}%
  \caption{\textbf{Single-photon electroluminescence from a cavity-coupled T centre. a}, Cavity reflectivity and photoluminescence excitation (PLE) spectrum of a single T centre with no applied magnetic field. A PLE linewidth full-width at half-maximum (FWHM) of 2.08(1) GHz is extracted from a Gaussian-Lorentzian product fit. The cavity is fit with a Lorentzian with a FWHM of 77(1)~GHz. The detuning $\Delta\nu$ reports the offset from $\nu_0=226.148$~THz. The inset shows an autocorrelation measurement under pulsed resonant optical excitation with a power of $\sim 30$~nW. \textbf{b}, Lifetime of the single T centre measured by pulsed resonant optical (red) and pulsed electrical (blue) excitation. The normalized counts collected in bins of 10~ns are fit to a single exponential function (dashed lines). The inset shows a simplified experimental setup. \textbf{c}, Background-corrected autocorrelation measurement under pulsed electrical excitation confirms electroluminescence from a single T centre. Error bars are $\pm1\sigma$.}
  \label{fig:single_photon}
\end{figure*}

\section{Results}
\subsection{Nanophotonic devices with integrated p-i-n diodes}
\label{sec:device_design}
Our hybrid optoelectronic devices integrate nanophotonic waveguides and cavities with lateral p-i-n junction diodes on an SOI chip (Fig.~\ref{fig:introduction}a). Two types of devices are presented in this work: tapered waveguide and cavity devices. These devices were fabricated on SOI using electron beam and optical lithography (Methods). A p- and n- implant density of $>10^{20}$~cm$^{-3}$ was chosen to achieve degenerate doping to form ohmic contacts and prevent carrier freeze-out at cryogenic temperatures. Fig.~\ref{fig:introduction}b shows the chemical structure of the T centre. Implantation parameters for T centre formation followed those reported by MacQuarrie et al.~\cite{macquarrie2021Generating}, with blanket carbon implantation of the entire chip. Masked hydrogen implantation was used to define the active T centre region in the devices and minimize passivation of the p- and n- dopants, which can impact the electrical characteristics of the diode~\cite{day2024Electrical}. A cross-section of the device layers for a p-i-n diode is shown in Fig.~\ref{fig:introduction}c. We measure current-voltage (I-V) characteristics of the diodes used in this work and find a turn-on voltage of 1.23~V for the tapered waveguide device, and 0.8~V for the cavity device, for a 1~µA current at 1.5~K. We find that the turn-on voltage increases at low temperatures with no limitation on current within our operating range, indicating that the implant density was sufficiently high to avoid carrier freeze-out (\hyperref[supmat:IV_curves]{Supplementary Section~\ref*{supmat:IV_curves}}).

The tapered waveguide and cavity devices both use single-mode waveguides and grating couplers (GCs) to couple light between the devices and a fibre array positioned above. The tapered waveguide devices can operate by transmission through the GCs, with one fibre above each. One GC of the device is designed to couple the zero-phonon line (ZPL) of the T centre, with the other designed for the phonon sideband (PSB)~\cite{deabreu2023Waveguideintegrated}. The tapered region expands the single-mode waveguide to support multiple modes and increases the coupled luminescence~\cite{deabreu2023Waveguideintegrated}. 

The cavity devices operate with a single GC designed for the ZPL and a single fibre, with an off-chip optical circulator to isolate the transmitted and reflected light. The cavity devices use zero-length 1D photonic crystal cavities~\cite{quan2011Deterministic}. A partially-etched waveguide crossing on each side of the cavity functions as an electrical connection between the metal contact and the doped waveguide region.

\begin{figure*}
  \makebox[\textwidth][c]{\includegraphics[width=180mm]{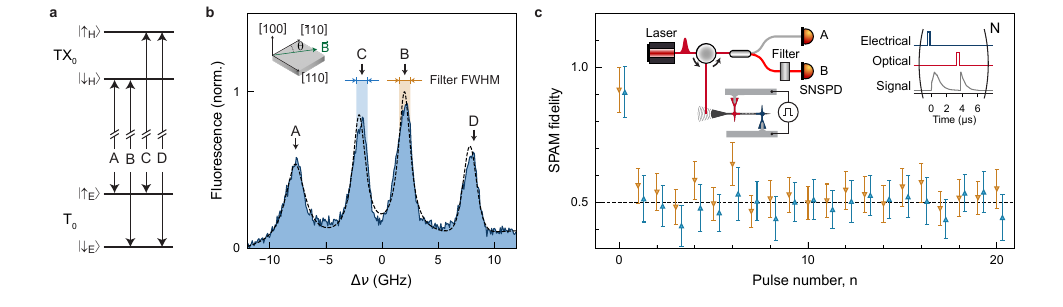}}%
  \caption{\textbf{Electrical spin initialization. a}, Spin-dependent optical transitions of the T centre under a magnetic field. \textbf{b}, PLE spectra collected using electrical/optical pulsing. $\textbf{B}=350$~mT, $\theta_{\mathrm{B}}\approx 13^{\circ}$ from $[\bar{1}10]$. The four peaks are fit with a sum of Gaussian-Lorentzian products (dashed line). The detuning $\Delta\nu$ reports the offset from $\nu_0=226.148$~THz. The FWHM of the spectral filter passband is shown when aligned to the B (orange) and C (blue) transitions. \textbf{c}, State preparation and measurement (SPAM) fidelity at pulse offset $n$ for preparing $\ket{\downarrow_\mathrm{E}}$ (orange inverted triangles) and $\ket{\uparrow_\mathrm{E}}$ (blue triangles), calculated from the correlation measurement ($\pm1\sigma$ error bars). The insets show a simplified experimental setup and the pulse sequence used for this experiment.}
  \label{fig:spin_init}
\end{figure*}

\subsection{Electroluminescence from a T centre ensemble}
We observe electroluminescence (EL) from a T centre ensemble in a tapered waveguide device at cryogenic temperatures (1.5~K), showing potential as an on-chip light source in the telecommunications O-band. The tapered design allows for excitation of a larger volume of T centres than a single-mode waveguide, a factor of $\sim 5$ for the device presented in this work. Taking the T centre generation density reported in Ref.~\cite{higginbottom2022Optical}, $1.7\times 10^{13}$~cm$^{-3}$, we estimate a lower bound of $\approx 800$ T centres in the active device area defined by the masked hydrogen implant. 

Fig.~\ref{fig:introduction}d shows the EL saturation curve of the tapered waveguide device up to a current of 250~μA where we observe peak count rates of 198(1)~kcps from the ZPL and 111(1)~kcps from the PSB GCs. We measure the EL spectrum (Fig.~\ref{fig:introduction}e) from a tapered waveguide device, choosing a constant forward-current of 5~μA to resolve the ZPL and PSB characteristics. With the two GCs we can observe characteristic features of the T centre including the ZPL transitions TX$_0$ and TX$_1$, as well as the PSB with a peak L$_1$ corresponding to the first local vibrational mode~\cite{bergeron2020SiliconIntegrated,deabreu2023Waveguideintegrated,irion1985defect}.

The inset in Fig.~\ref{fig:introduction}e shows the proposed mechanism for T centre EL, analogous to the photoluminescence (PL) process~\cite{davies1989optical}. Here, applying a sufficient forward bias to the diode injects free electrons and holes into the intrinsic region which can be captured by the T centre. From the bound exciton excited state, recombination produces a photon with some probability (the quantum radiative efficiency) which, based on the Debye-Waller factor, either generates a photon in the ZPL, or one or more phonon(s) and a photon in the PSB~\cite{bergeron2020SiliconIntegrated}. This process is non-resonant and excites a range of luminescent species in the sample. Light from other species is partially filtered by the GC before collection, where it acts as a bandpass filter. These results show that the tapered waveguide devices can be used as an on-chip light source operating in the telecommunications O-band, a T centre silicon light-emitting diode (LED).

\subsection{Single-photon electroluminescence}
\label{sec:el_spec}
We demonstrate single-photon EL from single cavity-coupled T centres. In the cavity devices, the p-i-n diode enables electrical excitation, with the emission rate enhanced by the Purcell effect. These devices allow single T centres to be controlled with both optical and electrical methods. We first measure the cavity spectrum to find devices with resonances close to the TX$_0$ ZPL. The data presented in this work are from one cavity device with a quality factor $Q=2{,}960(50)$ and a designed mode volume of $V\approx0.6(\lambda/n)^3$. We perform time-resolved photoluminescence excitation (PLE) spectroscopy, scanning the wavelength of a tuneable laser with pulsed excitation and measuring a luminescence decay transient. Performing PLE across the cavity range (with diode terminals floating), we observe emission from a T centre spectrally aligned with the cavity (Fig.~\ref{fig:single_photon}a). This PLE peak has a full-width at half maximum (FWHM) linewidth of 2.08(1)~GHz which is consistent with other reported single T centres in SOI~\cite{afzal2024Distributed,johnston2024Cavitycoupled,higginbottom2022Optical}.  To verify that this is a single T centre, we measure the second-order autocorrelation function $g^{(2)}(n)$ under pulsed resonant optical excitation (Fig.~\ref{fig:single_photon}a, inset). We extract a raw zero-delay coincidence rate $g^{(2)}_{\mathrm{raw}}(0)=0.36(1)<0.5$, confirming that the luminescence is predominantly from a single T centre.

Next, we investigate single-photon EL from the cavity-coupled T centre. Electrical excitation excites a range of defects naturally present in the silicon. This leads to a broad background luminescence which is spectrally filtered using a free-space bandpass filter and a fibre Fabry-P\'erot interferometer (Methods). We find that this device has an EL threshold voltage of $8$~V under forward bias, leading to $\sim 0.5$~mW dissipated over the device. This heating was found to cause the cavity resonance to blue-shift under prolonged bias. To reduce heating from the electrical excitation, a diode bias of 8~V is pulsed at 160~kHz with a pulse duration of 150~ns. Under this pulsed bias we did not observe any shift in cavity resonance.

We measure the transient decay from a single spectrally resolved T centre in pulsed operation, observing a lifetime of 570(1)~ns under electrical excitation compared to 431(1)~ns under resonant optical excitation (Fig.~\ref{fig:single_photon}b). The difference in lifetime between electrical and resonant optical excitation is consistent with the difference previously observed in T centre ensembles with above-bandgap and resonant optical excitation~\cite{deabreu2023Waveguideintegrated}. Comparing to the bulk lifetime of $\tau_0=0.94(1)$~µs~\cite{bergeron2020SiliconIntegrated}, we find a lifetime enhancement $\tau_0/\tau_{\mathrm{EL}} = 1.65(2)$ for electrical excitation and $\tau_0/\tau_{\mathrm{PLE}} = 2.18(2)$ for resonant optical excitation. As the centre is detuned from the cavity by $\Delta_{\mathrm{cav}}$, we can calculate the expected lifetime enhancement for the cavity by $\tau_0/\tau_{\mathrm{cav}}=P_t/[1+(2\Delta_{\mathrm{cav}}/\kappa)^2] + 1$. Where $\kappa=77(1)$~GHz, the cavity linewidth, $\Delta_{\mathrm{cav}}=15.65(3)$~GHz, and $P_t=(3/4\pi^2)(\lambda/n)^3 (Q/V) \eta_{\mathrm{DW}} \eta_{\mathrm{QE}}$ is the effective Purcell factor where $Q$ is the cavity quality factor, $V$ is the mode volume, $\eta_{\mathrm{DW}}=0.23$ is the Debye-Waller factor~\cite{bergeron2020SiliconIntegrated}, and $\eta_{\mathrm{QE}}=0.234$ is the quantum efficiency~\cite{johnston2024Cavitycoupled}. From this we find an expected lifetime enhancement of $\tau_0/\tau_{\mathrm{cav}} = 18.3$. Spatial and polarization misalignment may account for the difference.

We perform a pulsed correlation measurement under electrical excitation to extract the pulsed photon autocorrelation function, $g^{(2)}(n)$ (Fig.~\ref{fig:single_photon}c). We observe a raw zero-delay rate of $g^{(2)}_{\mathrm{raw}}(0)=0.358(7)$ and subtracting counts that can be attributed to background sources (Methods) we determine a background-subtracted coincidence rate of $g^{(2)}(0)=0.05(2)$ (\hyperref[supmat:autocorrelation]{Supplementary Section~\ref*{supmat:autocorrelation}}). We attribute the remaining zero-delay coincidence rate to luminescence from other defects---including other T centres. These results demonstrate electrically-triggered single-photon generation from a silicon colour centre for the first time. The same approach could be taken to isolate and electrically excite other defects of interest, such as single G and W centres in silicon~\cite{komza2024Indistinguishable,bao2007Point,buckley2017Allsilicon}.

\subsection{Spin initialization by electroluminescence heralding}
The T centre is not only a single-photon source, it also possesses a ground state electron spin qubit. Electrical injection of a spin-photon interface can be used to prepare the spin state by heralding on the spin-dependent EL. To evaluate spin initialization by EL heralding for the T centre, we first apply a magnetic field (\textbf{B}=350~mT, $\sim[\bar{1}10]$) to lift the electron spin state degeneracy. Under a magnetic field, TX$_0$ splits into four optically resolvable transitions: A--D (Fig.~\ref{fig:spin_init}a). A resonant laser can be used to optically address these four transitions, but exciting a single transition results in shelving population in the dark spin state. This hyperpolarization is typically lifted by using a second resonant laser to mix the electron spin~\cite{higginbottom2022Optical,deabreu2023Waveguideintegrated}. Here, we instead use electrical excitation to lift the hyperpolarization.

We measure the spectrum at field using a single pulsed laser alternating with an electrical `repump' pulse of 150~ns (Fig.~\ref{fig:spin_init}b). With this scheme we see that the electrical pulse mixes the electron spin state, allowing all four transitions to be resolved. We can infer that electrical excitation populates $\ket{\uparrow_\mathrm{H}}$ and $\ket{\downarrow_\mathrm{H}}$ with roughly equal probability, as shown by the fluorescence amplitudes in Fig.~\ref{fig:spin_init}b, with A and B corresponding to $\ket{\downarrow_\mathrm{H}}$ and C and D corresponding to $\ket{\uparrow_\mathrm{H}}$. The increased fluorescence of the higher energy transitions B and D is consistent with the spectral alignment of the transitions with the cavity resonance (Fig.~\ref{fig:single_photon}a). We fit the peak positions and calculate the hole-spin $g$-factor $g_h=1.18(1)$~\cite{higginbottom2022Optical}. We find that this value of $g_h$, under the applied field direction, is consistent with the [$(z1)$, $(z3)$] orientation subset which is also expected to have the best coupling to the transverse electric (TE) waveguide mode due to the $p_z$ like orbital~\cite{clear2024Optical,ivanov2022Effect}. We use the model from Ref.~\cite{clear2024Optical}, along with the measured external offset of $\theta_0=5(1)^{\circ}$, to estimate a magnetic field offset at the device of $\theta_{\mathrm{B}}\approx 13^{\circ}$ from $[\bar{1}10]$ (Fig.~\ref{fig:spin_init}b, inset). 

We now propose an electrically-triggered spin initialization scheme for the T centre with spectral heralding of the EL. In this scheme, we apply successive electrical pulses to excite the T centre, which then decays into either the $\ket{\uparrow_\mathrm{E}}$ or $\ket{\downarrow_\mathrm{E}}$ spin state by one of its four optical transitions. These transitions are optically resolved under a sufficient magnetic field, and by aligning a spectral filter to measure emission from one transition, detection of a transmitted photon heralds the final electron spin state. 

We measure the fidelity of this spectrally heralded spin state by splitting our collection into two paths - one with a spectral filter and one without. We introduce an optical `readout' pulse following each electrical pulse which is resonant with a spin-dependent optical transition. We perform a coincidence measurement between herald photons emitted from the electrical excitation, collected from the filtered path, and photons from subsequent readout optical excitation, collected from either path. We align the spectral filter to the B (C) transition, heralding preparation of the $\ket{\downarrow_\mathrm{E}}$ ($\ket{\uparrow_\mathrm{E}}$) spin state. Coincidence counts measured when the laser and the filter are aligned to the same spin-dependent optical transition, $c_{\mathrm{r}}(n)$, confirm the EL spin initialization. Conversely, coincidence counts measured when the laser and filter are aligned to different spin-dependent optical transitions, $c_{\mathrm{nr}}(n)$, correspond to infidelity in the spin preparation. The pulse number $n$ denotes the number of intermediate electrical-optical pulse cycles.

Fig.~\ref{fig:spin_init}c shows the state preparation and measurement (SPAM) fidelity given by  $\mathcal{F}(n)=c_{\mathrm{r}}(n)/(c_{\mathrm{nr}}(n) + c_{\mathrm{r}}(n))$. For $n>0$, $\mathcal{F} \approx 0.5$, demonstrating that the electrical excitation mixes the spin state. However, for $n=0$, where correlations are within a single pulse cycle, we find that we can prepare the ground state electron spin into $\ket{\downarrow_\mathrm{E}}$ ($\ket{\uparrow_\mathrm{E}}$) with a SPAM fidelity of $\mathcal{F}(0)=92(8)\%$ (91(9)\%), after background subtraction (\hyperref[supmat:fidelity_analysis]{Supplementary Section~\ref*{supmat:fidelity_analysis}}). This fidelity is limited by excitation of other T centres and defects in the device, which contribute to background luminescence (Fig.~\ref{fig:spin_init}b), as well as the time bin size (\hyperref[supmat:fidelity_analysis]{Supplementary Section~\ref*{supmat:fidelity_analysis}}). These results demonstrate that the spin can be initialized by spectrally-heralded, electrically-triggered single-photon emission. In future devices with fewer background defects and optimized device design, EL can be used to herald high-fidelity states (including superposition states) and distribute entanglement between centres~\cite{laccotripes2024Spinphoton}.

\section{Discussion}
In this work, we presented two types of nanophotonic devices with integrated p-i-n junction diodes for simultaneous optical and electrical control of silicon T centres. With the first waveguide-coupled device, we electrically excited a T centre ensemble, demonstrating an on-chip O-band LED in silicon. These compact on-chip light sources could be used to achieve high integration density in emerging classical optical computing technologies where cryogenic operation is desirable. 

With the second cavity-coupled device, we demonstrated, for the first time, electrically-injected single-photon emission from a silicon colour centre. Single-photon sources that operate in the telecommunications bands and are compatible with silicon photonics are a resource for quantum communication networks and all-optical quantum computers~\cite{knaut2024Entanglement, hofmann2012Heralded, kok2007Linear}. Our devices are waveguide and fibre coupled, allowing direct on-chip integration with silicon photonic switches, beamsplitters, and detectors, or efficient coupling to long-distance telecommunications networks~\cite{shekhar2024Roadmapping,yu2023Telecomband}.

To address the challenge of scaling solid-state quantum computers based on optically active spin defects, we proposed an electrical spin initialization scheme. Large-scale quantum systems may be constrained, in part, by the overhead required for spin initialization. This is considerable in traditional schemes that employ an optical switch network to address each device, as the networks required for commercially relevant quantum computing will number in the 100,000s of devices~\cite{vandam2023Using}. An alternative electrical spin initialization scheme can instead initialize many centres in parallel using on-chip filters and detectors~\cite{zheng2023OnChip,buckley2017Allsilicon}. Once initialized, further microwave operations could be used to create arbitrary quantum states, or a modulator can be used to herald a superposition state~\cite{moehring2007Entanglement}.

Using our device, we heralded the electron spin state of a single T centre by spectral filtering. This new mode of operation enhances the T centre's capabilities as an SPI, allowing for more efficient use in denser networks of devices, potentially lower power usage, and direct integration with cryo-CMOS circuitry. Future schemes may apply EL heralding to generate entanglement between remote SPIs~\cite{laccotripes2024Spinphoton}. 

These findings demonstrate the feasibility of silicon colour centres as silicon-integrated, electrically-injected light sources and single-photon sources. Such devices with G and W centres could be used as bright, electrically-injected single-photon sources for quantum networks and all-optical quantum computers~\cite{buckley2017Allsilicon,bao2007Point,komza2024Indistinguishable}. Similar diode devices have previously been applied to improve the optical coherence and to Stark tune solid-state emitters~\cite{anderson2019Electrical,candido2021Suppression,lukin2020Spectrally}. These devices enable equivalent investigations with the T centre~\cite{clear2024Optical}. The devices in this work could also be applied to emerging T centre-like silicon SPIs~\cite{xiong2024Computationally} with ground state electron spins, for electrically-injected spin initialization and remote entanglement. Ultimately, this work has shown that the T centre's spin-photon interface can be controlled electrically to enable highly parallel control, advancing the versatility of T centre quantum processors.

\bibliography{references}

\section{Methods}

\subsection{Device fabrication}
All fabrication, unless otherwise noted was performed by Photonic Inc. Devices were fabricated on commercial (100) SOI with a 220~nm device layer and 3~μm buried oxide layer. The chips were implanted over their full area with $^{12}$C at 38~keV and a fluence of $7\times 10^{12}$~cm$^{-2}$. The p-doped areas were patterned and implanted with B$^+$ at 30~keV and a fluence of $1.5\times 10^{15}$~cm$^{-2}$. The n-doped areas were patterned and implanted with P$^+$ at 62~keV and a fluence of $1.5\times 10^{15}$~cm$^{-2}$. Dopants were activated with a rapid thermal anneal at 1000~$^{\circ}\mathrm{C}$. The T centre regions were patterned and implanted with H$^+$ at 9~keV and a fluence of $7\times 10^{12}$~cm$^{-2}$. The chip underwent a rapid thermal anneal to activate the T centres as described in Ref.~\cite{macquarrie2021Generating}. This recipe and implant parameters for the carbon and hydrogen were chosen based on work from Refs.~\cite{macquarrie2021Generating,day2024Electrical}. Doping parameters for p- and n-type dopants were determined based on simulations with ‘Stopping Range of Ions in Matter’ (SRIM)~\cite{ziegler2010SRIM} targeting a dopant concentration of $\sim 10^{20}$~cm$^{-3}$ at a depth of 100~nm. All ion implantation was performed with a tilt of 7$^{\circ}$. Electron beam and optical lithography were used to transfer designed layers to the chips. A titanium-aluminum bilayer was used for electrical connections.

\subsection{Sample mounting and cryogenics}
The sample is mounted in a low-vibration closed-cycle cryostat (ICEoxford $^{\mathrm{DRY}}$ICE$^{\mathrm{1.5K}}$ 85~mm) operating at 1.5~K, unless otherwise noted. A custom printed circuit board (PCB) with a copper bottom clamps the sample into a copper mount with a recessed pocket, thermally anchoring the top face of the sample. The copper mount interfaces with a second copper plate, connecting to titanium piezoelectric nanopositioners (Attocube ANPx101 and ANPz102) through a gold-plated copper thermal link (Attocube ATC100/70) for three-axis movement.  The nanopositioners have resistive encoders which allow for closed-loop positioning of the sample under the optical fibre array with $\sim$2~µm precision over a range of 5~mm.

A six-port, 8.0$^{\circ}$ angle-polished, quartz v-groove fibre array (FiberTech Optica) is mounted in the cryostat above the chip to address devices with polarization-maintaining (PM) fibres (Thorlabs PM1300-XP). The fibres extend outside of the cryostat through a vacuum feedthrough, where they are individually terminated with FC/APC connectors.

Electrical connections to the chip are made by the same PCB which clamps the chip. It has two rows of 72 pins with 187~µm pitch, which are wire bonded with Al wire to the metallized pads on the chip, for a total of 144 connections. The connections are then routed to custom 80-pin flat-flex cables which connect to an interfacing PCB mounted on the plate above the chip. This interfacing PCB breaks out the connections further to six 24-pin PEEK circular connectors. The ICE Oxford cryostat has a 24-pin electrical feedthrough which is connected internally with a NbTi loom to the 24-pin PEEK connector near the interfacing PCB. 

\subsection{Device design}
Designs for the single-mode waveguide and bent waveguide follow from Ref.~\cite{deabreu2023Waveguideintegrated}. Design of the ZPL GC and the 1D photonic crystal cavities were provided by Photonic Inc. The PSB GC and waveguide crossing were designed for this work using Lumerical FDTD solver. The PSB GC was designed with linear apodization of the fill factor and grating period~\cite{marchetti2017Highefficiency}. The waveguide crossing has a parabolic taper and small footprint of 6$\times$6~µm, adapted from Ref.~\cite{bogaerts2007Lowloss} for operation at 1326~nm. Both the PSB GC and the waveguide crossing were optimized for a fixed shallow-etch depth of 60~nm.

The tapered waveguide device presented in this work has an intrinsic region width, $w_i=$~2.5~µm. The cavity device of this work has an intrinsic region width, $w_i=$~12.5~µm. The diodes in the tapered waveguide and cavity devices are formed perpendicular and parallel to the waveguide, respectively. The tapered waveguide devices have dedicated electrical connections, while the cavity devices are connected in parallel busses of 25.

\subsection{EL spectroscopy}
\label{sup:sec:EL_PL}
Electroluminescence (EL) is produced by applying a forward bias to the diode by a source measure unit (SMU) which can be configured for constant current or constant voltage operation (Keithley 2460). The current-voltage characteristics of the diodes were also measured using the SMU. 

The fluorescence spectrum is measured with a fibre-coupled spectrometer, consisting of a triple grating spectrograph (Princeton SpectraPro HRS-300) imaged by an ultra-low-noise LN-cooled InGaAs camera (Princeton NIRvana-LN). Gratings of 150, 600, and 1200 grooves per mm can be selected, for spectral resolutions of 0.8~nm, 0.2~nm, and 0.1~nm, respectively. The quantum efficiency of the camera is $>75\%$ at 1325~nm. Alternatively, the fluorescence can be directed to a fibre-coupled superconducting nanowire single-photon detector (SNSPD) with quantum efficiency $>70\%$ at 1325~nm (ID Quantique ID281). The single-photon detection events are logged by a multi-function time-tagger (ID Quantique ID900 Time Controller).

\subsection{PLE spectroscopy}
Photoluminescence excitation (PLE) spectra are measured by integrating the time-filtered decay transient from T centres under pulsed resonant excitation. The excitation wavelength is swept using a continuously tunable laser (Toptica CTL 1320) which is stabilized to a wavemeter (Bristol 871A-NIR) with a PID control loop. The output of the laser is maintained at 15~mW for mode-hop free tuning, with adjustable attenuation prior to the pulse amplification.

The tuneable laser is pulsed in two stages for a high extinction ratio. The first stage consists of a booster optical amplifier (BOA) (Thorlabs BOA1017P) with a high speed pulse driver (Aerodiode SOA-std) operating at a pulse current of 600~mA. The output is heavily attenuated prior to the second stage where an electro-optic amplitude modulator (Exail, MXER1300-LN-10) provides additional extinction. The EOM is controlled by an automatic bias controller (iXblue, MBC-DG-LAB) which re-biases the EOM before each scan, approximately every 30 minutes. The BOA and EOM are pulsed by a digital delay generator (Stanford Research System DG645) which is triggered by the IDQ ID900 time controller. This time-resolved excitation and detection allows for the lifetime of the emitter to be measured under resonant optical excitation. Devices are resonantly excited through the GC, with light coupled in/out of the fibre array by an optical circulator. Typical laser pulse power into the fibre array was  $\sim 75$~nW. Collected light is measured by the SNSPD and time-tagger as previously described. 

The PLE spectrum is recorded by sweeping the excitation wavelength and performing the pulsed measurement at each point. The excitation repetition rate is 500~kHz with a pulse duration is 100~ns, with a segmented  collection window with the first 900~ns consisting of `signal' and the following 450~ns consisting of `background'. The lifetime of the T centre was measured to be $\Gamma_o=431(1)$~ns under resonant optical excitation, resulting in the collection window containing nearly $90\%$ of the T centre emission. We subtract the `background' counts from the `signal' counts to obtain our background-corrected PLE spectra. This method compensates for: (1) leaked excitation laser and the wavelength dependence of the EOM (considerable during wide sweeps), (2) long-lived emission from other species in the waveguide, (3) detector dark counts.

Resolving the spin-dependent optical transitions with PLE spectroscopy under a magnetic field required pulsed electrical excitation prior to each optical pulse. In this work the electrical pulse duration was 150~ns with a total pulse period of 6.25~µs.

\subsection{Pulsed electrical excitation}
Electrical pulses were generated using a digital delay generator (Stanford Research System DG645) to switch a power supply in constant voltage mode (Keysight EDU3611A) through an N-channel MOSFET (IRF130). The diodes of the cavity devices require a forward voltage above 8~V for EL from the device shown in this work. The DC wiring in the cryostat causes the devices to act as a large capacitive load which requires a source with low output impedance for fast switching. To improve the switching speed, a 50~Ω, 2~W resistor is placed outside of the cryostat, in parallel with the diodes. Despite this, the slow rise time of the MOSFET and electrical reflections result in an applied 150~ns pulse being stretched to approximately 750~ns.

Pulsed electrical excitation, when used for single-photon EL and spin initialization, requires spectral filtering to remove emission from other species present in the device. The filtering consists of a free-space 1329$\pm$0.5~nm bandpass filter (Iridian) which is tilted to shift the passband to 1325.5$\pm$0.5~nm. This is followed by a fibre Fabry-P\'erot interferometer (FFPI) with a bandwidth of $0.985$~GHz and a free spectral range of $102.4$~GHz (LUNA, FFP-I). The combination of these two filters allows for a single FFPI cavity mode to be transmitted, with resonance tuning by temperature control of the FFPI (Thorlabs TED200C). A schematic of the setup can be found in \hyperref[supmat:experimental_setup]{Supplementary Section~\ref*{supmat:experimental_setup}}.

\subsection{Photon autocorrelation}
We measure the autocorrelation function using correlations between two detectors in a Hanbury-Brown-Twiss configuration. Optical autocorrelation measurements use photons detected during a collection window similar to that used for PLE spectroscopy. The optical autocorrelation measurement presented in this work was captured at 2.5~K with a power of 30~nW, a pulse duration of 750~ns, a pulse period of 3.568~µs, and a collection window of 1.5~µs. Electroluminescence (EL) autocorrelation measurements use all detected photons, with background subtraction performed based on the measured lifetime of the emitter (\hyperref[supmat:autocorrelation]{Supplementary Section~\ref*{supmat:autocorrelation}}). The EL autocorrelation measurement presented in this work was captured at 1.5~K with a voltage of 8~V, a pulse duration of 150~ns, a pulse period of 4~µs, and a collection window of 4~µs.

Raw measurements of the coincidence rate for a given pulse delay $n$ are denoted as $g^{(2)}_{\mathrm{raw}}(n)$ and are calculated using the integrated coincidence counts in the pulse period with the normalization factor $\mathcal{N}=N_1 N_2 \theta T$, where $N_1$, $N_2$ are the detector count rates, $\theta$ is pulse repetition period, and $T$ is the total acquisition time. This corresponds to the integrated area of a Poisson distribution~\cite{beveratos2002Room}.

Background corrected measurements are obtained by fitting a series of positive and negative time-delayed correlations from 20 pulses, for a total of 41 exponential peaks. We extract the background level by simultaneously measuring the emitter lifetime $\tau_c$ and using this as a fixed parameter~\cite{laferriere2023PositionControlled}. In this manner we can remove the background contributions from other defects and dark counts. We collect timestamps into $d=40$~ns bins and model the autocorrelation function as a function of time delay $\tau$

\begin{equation}
    \begin{aligned}
        g^{(2)}(\tau)&=I_0 + A_0 e^{-|\tau|/\tau_c} \\
        &+ \sum_{n=1}^{N}(A_{n}e^{-|\tau-n\theta|/\tau_c} + A_{(-n)}e^{-|\tau+n\theta|/\tau_c})
    \end{aligned}
\end{equation}

where $A_0$ is the height of the peak at $\tau=0$ while $A_{n}$ are the heights of the full-sized peaks at pulse delay $n$. The amplitudes $A_n$ are then used to calculate the second-order correlation function as a function of pulse delay 
\begin{equation}
    g^{(2)}(n)=\dfrac{2A_n\tau_c}{\mathcal{N}-I_0\theta/d}
\end{equation}

Further details about the autocorrelation fitting procedure can be found in \hyperref[supmat:autocorrelation]{Supplementary Section~\ref*{supmat:autocorrelation}}.

\subsection{Spin initialization measurements}
We calculate the correlations between photons from electrical pulses and optical pulses to test the spin initialization by EL heralding scheme. We analyze cross-correlations between two collection windows of width $t_{\mathrm{ce}}$ and $t_{\mathrm{co}}$, offset from the end of the electrical and optical excitation pulses by delay $t_{\mathrm{e}}$ and $t_{\mathrm{o}}$, respectively. The window width and offsets can be tuned to optimize the SPAM fidelity, in this work the selected values for preparing $\ket{\downarrow_\mathrm{E}}$ ($\ket{\uparrow_\mathrm{E}}$) are $t_{\mathrm{ce}}=500$~ns (500~ns), $t_{\mathrm{co}}=411$~ns (283~ns), $t_{\mathrm{e}}=418$~ns (515~ns), and $t_{\mathrm{o}}=65$~ns (65~ns). The coincidence counts are background subtracted before calculating the fidelity by measuring the contributions from other defects and leaked excitation with the laser and filter detuned between the A and C, C and B, and B and D transitions. Including those aligned to B and C, this results in a total of 25 laser/filter combinations. The experiment is run for a fixed filter position, cycling through the five laser positions. The filter position is then changed by setting the laser wavelength to the desired position, and maximizing transmission by adjusting the FFPI temperature. A transmission spectrum is captured to confirm the position. 

Background counts are made up of the following: (1) Emission from other defects excited during the EL pulse, (2) Leaked EL from other T centre transitions, (3) Emission from other defects excited during the optical pulse, (4) Cross-excitation from other T centre transitions. A linear fit to the background measurements is used to calculate the contribution from other defects when aligned to the B and C transitions. The measured PLE spectrum is used to calculate contributions from cross-excitation and filter overlap. Without background subtraction, the raw SPAM fidelity for preparing the ground state electron spin into $\ket{\downarrow_\mathrm{E}}$ ($\ket{\uparrow_\mathrm{E}}$) was $\mathcal{F_{\mathrm{raw}}}(0)=87(8)\%$ (86(9)\%). Further details can be found in \hyperref[supmat:fidelity_analysis]{Supplementary Section~\ref*{supmat:fidelity_analysis}}.

\section*{Author Contributions}
M.D., S.S. and D.B.H designed the experiment. M.D. and F.A. designed the samples used in the study. I.M. and N.J. fabricated the samples. C.C. assisted in conception of the experiment and device design. M.D., C.B., S.A.M., and M.G. built the measurement apparatus. M.D., S.A.M., and E.H. measured diode characteristics. M.D. measured electroluminescence, lifetime, photon correlations, and spin initialization SPAM fidelity.  S.A.M., C.D., and M.L.W.T. advised on design and analysis. All authors participated in preparation of the manuscript.\vspace{0.2cm}

\section*{Acknowledgements}
The authors would like to thank the Integrated Photonics team at Photonic Inc. for their contributions to the design and fabrication of the silicon chip presented in this work.

This work was supported by the Natural Sciences and Engineering Research Council of Canada (NSERC), the New Frontiers in Research Fund (NFRF), the Canada Research Chairs program (CRC), the Canada Foundation for Innovation (CFI), the B.C. Knowledge Development Fund (BCKDF), the Quantum Information Science program at the Canadian Institute for Advanced Research (CIFAR), and Photonic Inc.

S.A.M. acknowledges support from NSERC (PDF - 587831 - 2024). S.S. is supported by the Arthur B. McDonald Fellowship.

\section*{Competing Interests}
M.D., C.B., S.A.M., C.C., M.G., I.M., F.A., C.D., N.J., M.L.W.T., S.S, and D.B.H are current or recent employees of and/or have a financial interest in Photonic Inc., a quantum technology company. E.H. declares no competing interests.

\section*{Data Availability}
The data that support this work are available from the corresponding author upon reasonable request.

\onecolumngrid
\vfill
\clearpage

\renewcommand{\thefigure}{S\arabic{figure}}
\setcounter{figure}{0}
\setcounter{section}{0}
\setcounter{subsection}{0}
\vspace*{3cm}
\pagenumbering{gobble}
\begin{center}
\begin{minipage}{.8\textwidth}
\centering
\Huge{Supplementary information for ``Electrically-triggered spin-photon devices in silicon''}
\end{minipage} 

\vspace{1.5cm}

\begin{minipage}{.9\textwidth}
\centering
\Large M. Dobinson$^{1,2}$, C. Bowness$^{1,2}$, S. A. Meynell$^{1,2}$, C. Chartrand$^{1,2}$, E. Hoffmann$^{3}$, M. Gascoine$^{1,2}$, I. MacGilp$^{2}$, F. Afzal$^{2}$, C. Dangel$^{2}$, N. Jahed$^{2}$, M. L. W. Thewalt$^{1,2}$, S. Simmons$^{1,2}$, D. B. Higginbottom$^{1,2\dagger}$
\end{minipage} 

\vspace{1.5cm}

\begin{minipage}{.7\textwidth}
\centering
\large$^1$Department of Physics, Simon Fraser University, Burnaby, British Columbia, Canada \vspace{0.2cm}

$^2$Photonic Inc., Coquitlam, British Columbia, Canada \vspace{0.2cm}

$^3$Département de physique, Université Paris-Saclay, 91190 Gif-sur-Yvette, France\vspace{1cm}

$^{\dagger}$Corresponding author. Email: \href{mailto:daniel_higginbottom@sfu.ca}{daniel\_higginbottom@sfu.ca} 
\end{minipage}

\end{center}

\normalsize
\clearpage

\section{Experimental setup}
\begin{figure*}[ht!]
  \makebox[\textwidth][c]{\includegraphics[width=180mm]{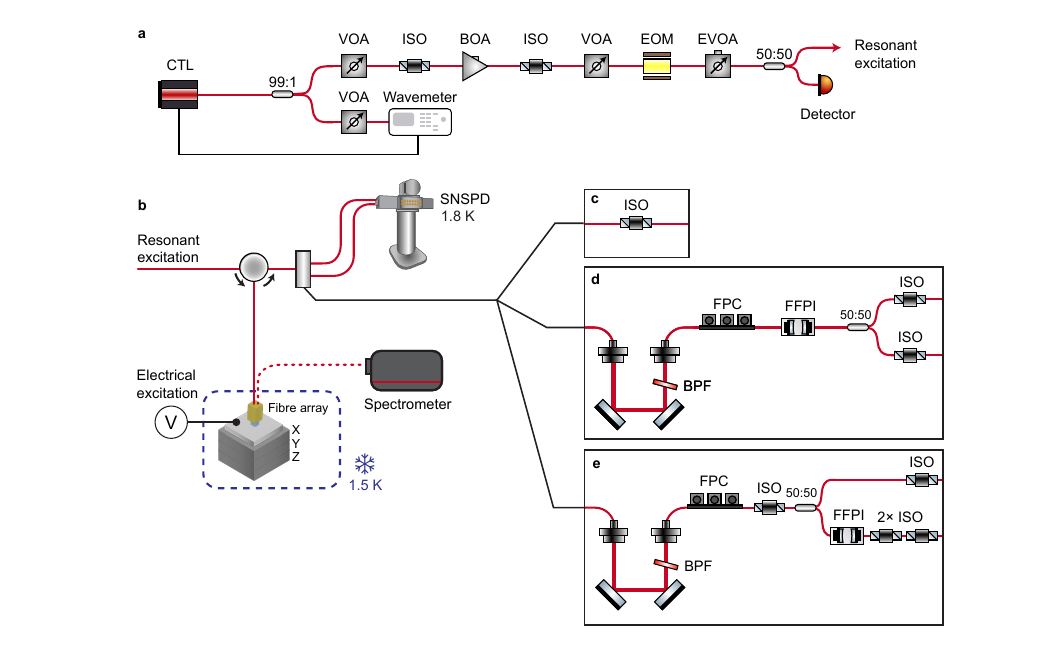}}%
  \caption{\textbf{Experimental setup. a}, Resonant optical excitation. \textbf{b}, Chip excitation and detection. The fibre block has six fibres which can be connected for different experiments. \textbf{c}, Setup for PLE. \textbf{d}, Setup for EL HBT measurements. \textbf{e}, Setup for spin initialization measurements. Additional isolators are used to minimize reflections from the FFPI and SNSPDs. BPF: bandpass filter, BOA: booster optical amplifier, EOM: electro-optic modulator, EVOA: electronically-variable optical attenuator, FFPI: fibre Fabry-P\'erot interferometer, FPC: fibre polarization controller, ISO: isolator, VOA: variable optical attenuator.}
  \label{fig:experiment_diagram}
\end{figure*}

\pagenumbering{arabic}
\label{supmat:experimental_setup}
In this section we provide a detailed experimental schematic for the experiments in the main text. Fig.~\ref{fig:experiment_diagram}a,b show the excitation and detection paths. Fig.~\ref{fig:experiment_diagram}c-e break out the optical components which are placed before the SNSPD for PLE spectroscopy, EL autocorrelation, and spin initialization measurements, respectively.

\clearpage
\section{IV characteristics of the p-i-n diodes}
\label{supmat:IV_curves}
Here we present the current-voltage (IV) characteristics of the p-i-n diodes from the tapered waveguide and cavity devices presented in the main text. The tapered waveguide device was measured individually while the cavity device is connected to a parallel bus of 25 devices. The IV characteristics were measured using a source measure unit (SMU) to sweep the applied voltage (Keithley 2460). The voltage was swept from 0~V until a set current was reached, 100~μA for the tapered waveguide devices and 2.5~mA for the cavity devices. To avoid hysteresis effects from heating, the voltage was returned to 0~V for 5 seconds before sweeping to -10~V. The IV curves were fit to extract the resistance under forward and reverse bias.

\begin{figure*}[h]
  \makebox[\textwidth][c]{\includegraphics{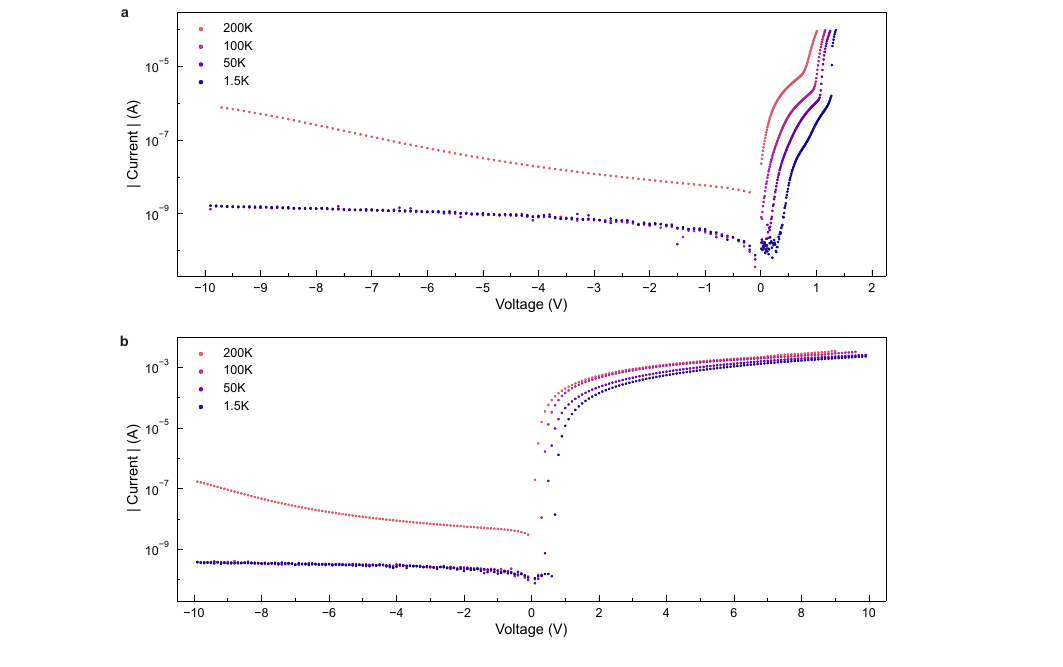}}%
  \caption{\textbf{IV curves of the two p-i-n diodes used in the main work. a,} Tapered waveguide device with forward and reverse resistances at 1.5~K of 990~Ω and 7.2~GΩ, respectively. \textbf{b,} Cavity devices (bus of 25 devices) with forward and reverse resistances at 1.5~K of 3,200~Ω and 58~GΩ, respectively.}
  \label{fig:IV}
\end{figure*}

\clearpage

\section{Autocorrelation analysis}
\label{supmat:autocorrelation}
The EL autocorrelation measurements presented in the main text have been background-corrected using the procedure detailed in the Methods. The total count rate collected from the electrically excited single T centre during this measurement was 209 cps, averaged over the 4~µs cycle. With this count rate, the broad luminescent background and detector dark counts are a significant contribution. The background fitting follows the procedure described in Ref.~\cite{laferriere2023PositionControlled}. 

\begin{figure*}[h]
  \makebox[\textwidth][c]{\includegraphics{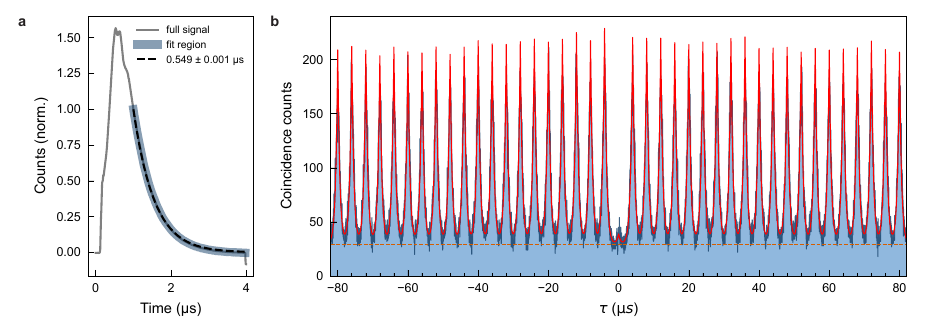}}%
  \caption{\textbf{Autocorrelation measurement with background subtraction. a}, Lifetime fitting. \textbf{b}, Uncropped fit to all 41 exponential peaks using 40~ns bins.}
  \label{fig:g2_backsub}
\end{figure*}

Fig.~\ref{fig:g2_backsub}a shows the lifetime measured during the autocorrelation measurement. Due to limitations of the setup, we find that the electrical pulse is stretched to $\sim 750$~ns. After the pulse, there is a period of fast decay from other excited defects transmit through the spectral filter. We wait 1~µs after the electrical pulse to measure the lifetime of the T centre. We use the measured lifetime as a fixed parameter to fit the background level, which removes contributions from other emitters with significantly longer lifetimes. As this measurement also spectrally filters the emission, the majority of the luminescence is from the single T centre, and the fixed lifetime in the fit acts as a temporal filter. Fig.~\ref{fig:g2_backsub}b shows the full fit to 41 exponential peaks with 40~ns bins.

\begin{figure*}[h]
  \makebox[\textwidth][c]{\includegraphics{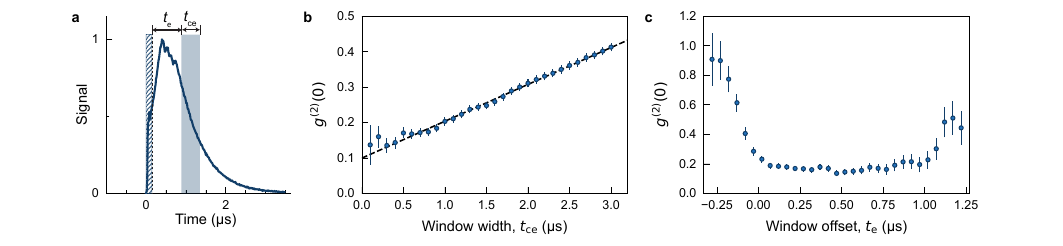}}%
  \caption{\textbf{Autocorrelation measurement with temporal filtering. a}, Window position for EL pulse with the hashed and shaded regions showing the excitation pulse and collection window, respectively. \textbf{b}, Varying the window width $t_{\mathrm{ce}}$. \textbf{c}, Varying the window offset $t_{\mathrm{e}}$.}
  \label{fig:g2_timefilt}
\end{figure*}

We can also compare this method to temporally filtering the raw timestamp data without background subtraction. We can limit the analysis to photons detected in a collection window of a specific width $t_{\mathrm{ce}}$ which is offset from the electrical pulse by a time $t_{\mathrm{e}}$. Fig.~\ref{fig:g2_timefilt}b shows that as we vary $t_{\mathrm{ce}}$ from 0.1~µs to 3.0~µs, with $t_{\mathrm{e}}=0.47$~µs, $g^{(2)}(0)$ increases, which is consistent with background counts being a significant factor. With a linear fit we find an intercept at $t_{\mathrm{ce}}=0$ of $g^{(2)}(0) = 0.010(1)$. We vary the offset $t_{\mathrm{e}}$ in Fig.~\ref{fig:g2_timefilt}c with a fixed width of $t_{\mathrm{ce}}=0.3$~µs and find that the $g^{(2)}(0)$ is minimized for $t_{\mathrm{e}}=0.47$~µs.

\newpage
\section{SPAM fidelity analysis}
\label{supmat:fidelity_analysis}
The following sections give additional details on two main parts to the SPAM fidelity analysis: selecting the window parameters and background subtraction.

\subsection{Windowing}
\begin{figure*}[h]
  \makebox[\textwidth][c]{\includegraphics{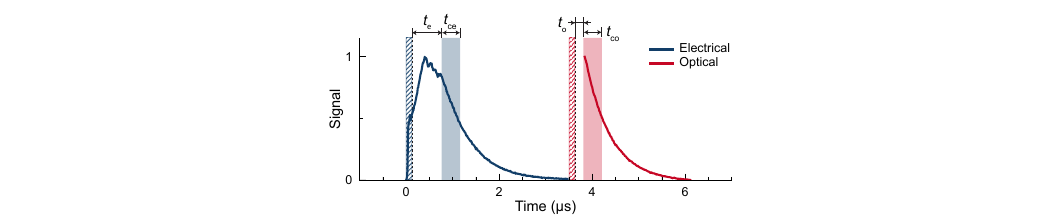}}%
  \caption{\textbf{Collection windows for fidelity analysis.} Signal collected from spin initialization measurements. The excitation pulses are shown as hashed regions and the collection windows are shaded for the electrical (blue) and optical (red) pulses. Collection window offsets, $t_{\mathrm{e}}, t_{\mathrm{o}}$ and widths $t_{\mathrm{ce}}, t_{\mathrm{co}}$ are shown.}
  \label{fig:fidelity_window}
\end{figure*}

The spin initialization measurement captures raw timestamps for photon clicks from electrical and optical excitation. When analyzing the data we can choose a window in which we perform the cross-correlation. If the spin initialization scheme were to be used in a real implementation, an appropriate window can be chosen to optimize the fidelity of the system. Fig.~\ref{fig:fidelity_window} shows the two parameters which can be controlled: the window offset and width.

The window offset is the delay from end of the excitation pulse to the start of the cross-correlation window. This is denoted $t_{\mathrm{e}}$ for the electrical pulse and $t_{\mathrm{o}}$ for the optical pulse. Under electrical excitation there are a number of other species in the device which are excited, and with a window offset we reduce the contribution from species with short lifetimes. Resonant optical excitation is more selective, but some delay is needed to avoid capturing leaked excitation light and reflections which follow the optical pulse. 

The window width is measured from the end of the window offset and can be selected to maximize the signal-to-noise ratio (SNR). At later times, the signal is dominated by dark counts, and by reducing the duration we can improve the fidelity. This width is denoted $t_{\mathrm{ce}}$ for the electrical pulse and $t_{\mathrm{co}}$ for the optical pulse.

\begin{figure*}[ht!]
  \makebox[\textwidth][c]{\includegraphics[width=180mm]{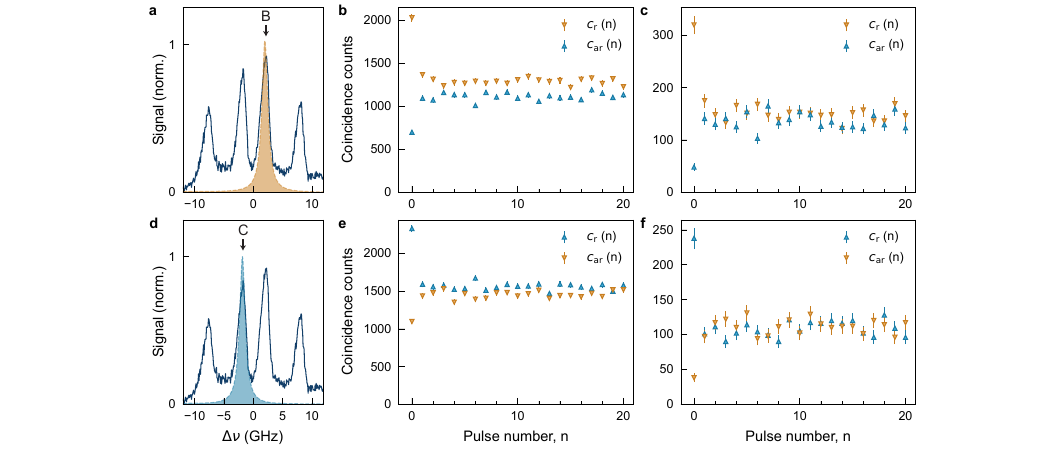}}%
  \caption{\textbf{Coincidence counts without background subtraction. a--c}, Filter aligned to B transition, showing spectrum (a), coincidence counts before (b) and after (c) windowing. \textbf{d--f},  Filter aligned to C transition, showing spectrum (d), coincidence counts before (e) and after (f) windowing.}
  \label{fig:spin_init_windowed}
\end{figure*}

As described in the main text, the cross-correlation measurement results in coincidence counts $c_{\mathrm{r}}(n)$ when the filter and laser are aligned to the same transition, and $c_{\mathrm{nr}}(n)$ when the filter and laser are aligned to different transitions. The effects of windowing are shown in Fig.~\ref{fig:spin_init_windowed}, where the coincidence counts for each case before and after windowing are plotted.

\begin{figure}[ht!]
  \makebox[\textwidth][c]{\includegraphics[width=180mm]{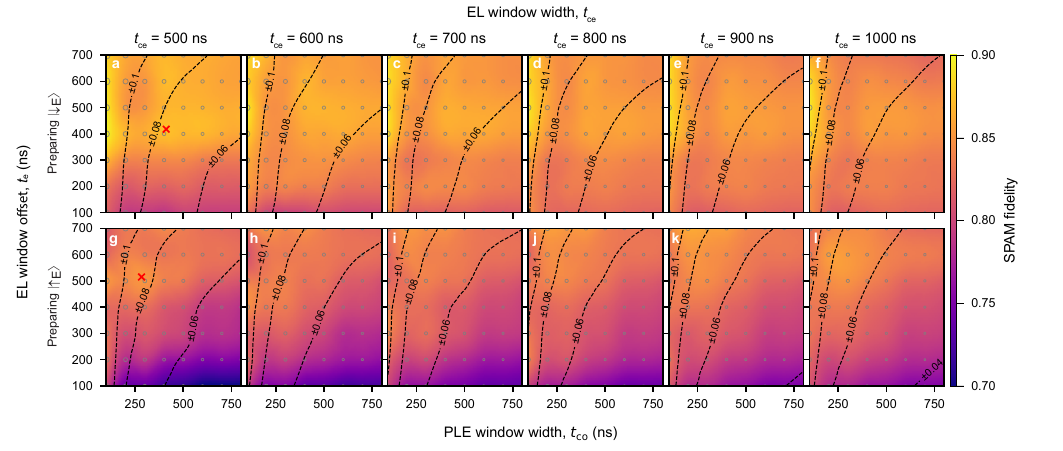}}%
  \caption{\textbf{SPAM fidelity collection window sweep, without background subtraction.} Two collection windows are used to filter counts from electrical and optical excitation. The grey circles indicate the calculated points, with the size of the circle related to the uncertainty. Clough-Tocher interpolation is used to fill between the calculated points. The dashed black contours show the fidelity uncertainty. The red crosses mark the window parameters used in the main text. \textbf{a--f}, Preparing $\ket{\downarrow \mathrm{E}}$. \textbf{g--i}, Preparing $\ket{\uparrow \mathrm{E}}$.}
  \label{fig:spin_init_window}
\end{figure}

The collection window can also be optimized to improve the fidelity. We can post-process the timestamp data to find the optimal window offset and width for our device. Fig.~\ref{fig:spin_init_window} shows the fidelity calculated for combinations of $t_{\mathrm{e}}, t_{{\mathrm{ce}}},$ and $t_{{\mathrm{co}}}$, with fixed $t_{\mathrm{o}}=65$~ns.

\clearpage
\subsection{Background subtraction}
Background subtraction aims to separate the contribution from other emitters in the device and dark counts. This yields the signal that can be attributed to the emitter of interest and indicates the performance of future devices with targeted defect formation. The background data is captured alongside the main experiment, in which we use five filter positions and five optical excitation wavelengths for a total of 25 combinations. The background from the electrical and optical collection windows are calculated separately. The filter transmission spectrum is fit with a normalized Gaussian-Lorentzian product, $\gamma_{F,i}(\nu)$, for each position. We use the fitted PLE spectrum from the main text as the upper-bound of excitation and emission probability (as stronger transitions are saturated before weaker ones in PLE). The fitted function $\mathcal{T}(\nu)$ is a sum of four Gaussian-Lorentzian products, $\gamma_i(\nu)$ for transition $i$. We can write the full fit as:

\begin{equation}
    \mathcal{T}(\nu)=\sum_{i\in\{A,B,C,D\}} \gamma_i(\nu)
\end{equation}

\begin{figure*}[ht!]
  \makebox[\textwidth][c]{\includegraphics[width=180mm]{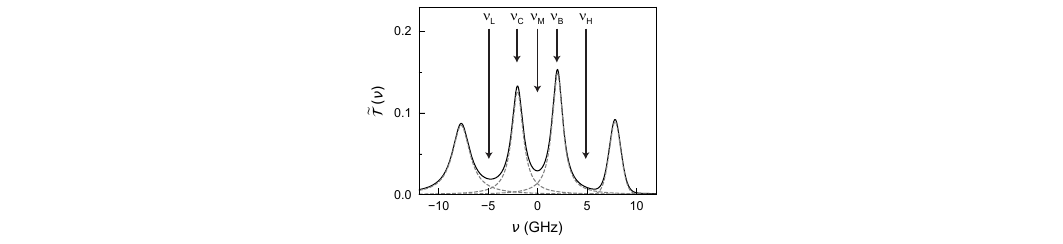}}%
  \caption{\textbf{Model of PLE spectrum.} Fitted spectrum showing how $\mathcal{T}(\nu)$ is normalized to its area.}
  \label{fig:ple_spec_model}
\end{figure*}

To determine the background signal for the electrical window, we extract the background luminescence by measuring the emission with the filter detuned from the transitions of interest, at three positions denoted $L, M, H$. These positions are located on either side of and in between the B and C transitions. We collect an average count rate $c_{i}(\nu_j)$ for filter position $i\in\{L, M, H\}$ at excitation frequency $\nu_j$. Although detuned, the filter will still collect some luminescence from the T centre, so we calculate the overlap of the T centre emission with the filter (normalized to the total area) and subtract it from the measured count rate

\begin{equation}
    f^{(E)}_{j}(\nu_i) = \Big(1-\dfrac{\int{\gamma_{F,i}(\nu)\mathcal{T}(\nu)~\,d\nu}}{\int{\mathcal{T}(\nu)~\,d\nu}}\Big)c_{i}(\nu_j)
\end{equation}

where $f^{(E)}_{j}(\nu_i)$ are the resulting EL background counts for the filter position $i\in\{L, M, H\}$ which are fit for filter position $j\in\{B,C\}$. We perform a linear fit to these three positions to estimate the background when the filter is at the B and C transitions, denoted $b_{B}^{(E)}$ and $b_{C}^{(E)}$.

We follow a similar procedure for the optical window, in which we excite at three positions denoted $\nu_L, \nu_M, \nu_H$ with the filter fixed at transition $i\in\{B,C\}$. We collect the average count rate $c_i(\nu_j)$ for filter position $i\in\{B,C\}$ at excitation frequency $\nu_j$. To find the background we must subtract the proportion of emission from optical excitation of the T centre. We can calculate the probability of exciting transition $i$ with the laser at $\nu_j$

\begin{equation}
    p^{(O)}_i(\nu_j) = \dfrac{\gamma_i(\nu_j)}{\int{\mathcal{T}(\nu)\,d\nu}}
\end{equation}

The optical collection window is made up of two counts collected over two paths, one with the filter and one without. The fraction of counts collected from the unfiltered path is denoted $\eta=0.693(1)$. We can now sum over each transition, and with the probability of exciting it with our laser at $\nu_j\in\{\nu_L, \nu_M, \nu_H\}$, calculate the fraction of emission that is expected to be included in the background count rate $c_i(\nu_j)$ for filter position $i$

\begin{equation}
    f^{(O)}_i(\nu_j) = \Big(1-\sum_{n\in\{A,B,C,D\}}p^{(O)}_n(\nu_j)\dfrac{\int{(\eta\gamma_n(\nu) + (1-\eta)\gamma_n(\nu)\gamma_{F,i}(\nu))\,d\nu}}{\int{\mathcal{T}(\nu)\,d\nu}}\Big)c_i(\nu_j)
\end{equation}

where $f^{(O)}_{i}(\nu_j)$ are the resulting PLE background counts for the filter position $i\in\{B,C\}$ with excitation at $\nu_j\in\{\nu_L, \nu_M, \nu_H\}$ which are fit for excitation at $\nu_j\in\{\nu_B,\nu_C\}$. We perform a linear fit to estimate the background in the optical collection window with filter and excitation combinations at the B and C transitions, denoted $b_{ij}^{(O)}$ for $i,j\in\{B,C\}$.

After calculating the background count rate for both the electrical and optical collection windows, we can calculate the background counts per pulse. With the filter aligned to transition $i\in\{B,C\}$ and excitation of transition $j\in\{B,C\}$ we find 

\begin{equation}
    \mathcal{B}_{ij}=b_{i}^{(E)}b_{ij}^{(O)}\theta T
\end{equation}

where $\theta$ is the full pulse duration and $T$ is the total acquisition time. We can subtract $\mathcal{B}_{ij}$ from the correlation counts $c_{\mathrm{r}}(n)$ and $c_{\mathrm{nr}}(n)$ to yield the final background subtracted values. 

\clearpage
\end{document}